\documentclass[utf8]{FrontiersinHarvard} 

\usepackage{url,hyperref,lineno,microtype,subcaption,rotating,lscape,graphicx,ragged2e}
\usepackage[onehalfspacing]{setspace}
\usepackage{multirow}

\def\keyFont{\fontsize{8}{11}\helveticabold }
\def\firstAuthorLast{Sina Labbaf {et~al.}} 
\def\Authors{Sina Labbaf\,$^{1,*}$, Mahyar Abbasian\,$^{1}$, Iman Azimi\,$^{2}$, Nikil Dutt\,$^{1,2}$, and Amir M. Rahmani\,$^{1,2,3}$ }
\def\Address{$^{1}$Department of Computer Science, University of California, Irvine, CA, USA \\
$^{2}$Institute for Future Health, University of California, Irvine, CA, USA \\
$^{3}$School of Nursing, University of California, Irvine, CA, USA
}
\def\corrAuthor{Sina Labbaf, 3069 Donald Bren School of Information and Computer Sciences, University of California, Irvine}
\def\corrEmail{slabbaf@uci.edu}

\begin{document}
\onecolumn
\firstpage{1}

\title[ZotCare]{ZotCare: A Flexible, Personalizable, and Affordable mHealth Service Provider} 

\author[\firstAuthorLast ]{\Authors} 
\address{} 
\correspondance{} 

\extraAuth{}

\maketitle

\begin{abstract}

\section{}
The proliferation of Internet-connected health devices and the widespread availability of mobile connectivity have resulted in a wealth of reliable digital health data and the potential for delivering just-in-time interventions. 
However, leveraging these opportunities for health research requires the development and deployment of mobile health (mHealth) applications, which present significant technical challenges for researchers. 
While existing mHealth solutions have made progress in addressing some of these challenges, they often fall short in terms of time-to-use, affordability, and flexibility for personalization and adaptation.
ZotCare aims to address these limitations by offering ready-to-use and flexible services, providing researchers with an accessible, cost-effective, and adaptable solution for their mHealth studies. 
This article focuses on ZotCare's service orchestration and highlights its capabilities in creating a programmable environment for mHealth research. 
Additionally, we showcase several successful research use cases that have utilized ZotCare, both in the past and in ongoing projects.
Furthermore, we provide resources and information for researchers who are considering ZotCare as their mHealth research solution. 

\tiny
 \keyFont{ \section{Keywords:} mobile health, mHealth solution, health cybernetics, digital health services, wearable internet-of-things} 
\end{abstract}

\section{Introduction} 

The widespread adoption of smartphones, wearable technologies, and other Internet-connected health devices has led to the availability of reliable digital health data streams \citep{health-iot-opportunities}. These devices and applications have played a significant role in various domains, such as improving lifestyles, achieving fitness goals, monitoring high-risk populations, and enhancing productivity \citep{medical-iot-devices-review}. Many vendors now offer access to the data streams generated by their products, opening up new opportunities for researchers to explore ubiquitous remote monitoring by leveraging different health data streams \citep{samsung-device-sensors,wearos-dev,garmin-dev,whitings-dev,fitbit-dev,oura-dev}. For instance, studies such as \cite{finland-pregnant-sleep} and \cite{dementia-garmin} have utilized Garmin smartwatches \cite{garmin-dev} to longitudinally monitor maternal sleep and dementia patients' caregivers, respectively.
Furthermore, the rise in mobile Internet connectivity \citep{mobile-connectivity} has provided researchers with the ability to promptly interact with participants, facilitating the collection of supplementary information for data modeling or the delivery of interventions within minutes or seconds.
By capitalizing on these two opportunities, researchers can not only collect accurate health data streams but also process the information, engage with participants, and implement necessary interventions.

\begin{figure*}[ht]
\begin{center}
\includegraphics[width=\textwidth]{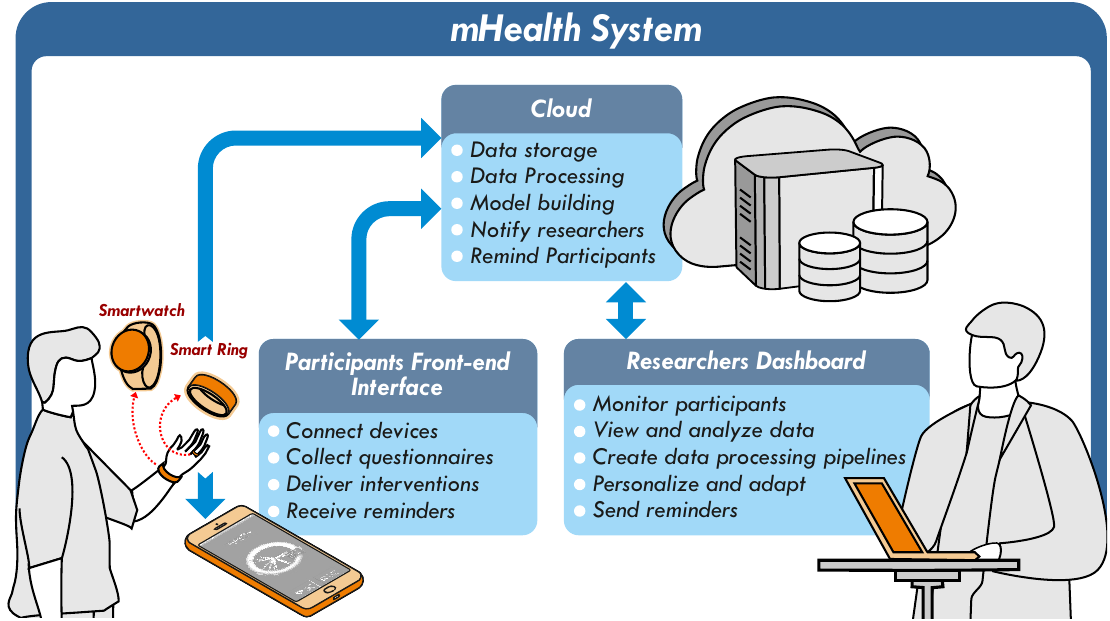}
\end{center}
\caption{An overview of a mHealth system}
\label{fig:mhealth-system}
\end{figure*}

For health researchers, leveraging these opportunities necessitates developing and deploying mobile health (mHealth) applications.
These applications perform tasks such as collecting health data streams, processing the data, invoking actions, and receiving feedback.
Figure \ref{fig:mhealth-system} outlines a typical mHealth system composed of three critical components: the central cloud server and separate interfaces for researchers/clinicians and participants.
The cloud server forms the foundation for data storage, model building, and action invoking aimed at participants and researchers, all while ensuring the preservation of data integrity, security, and participant privacy. 
The participant interface, another critical component, necessitates real-time interaction capabilities with participants, as well as mechanisms for subjective and objective data collection. 
Conversely, the researcher dashboard should be furnished with data analysis and monitoring tools essential for executing a mHealth study.
Each component operates within a distinct segment of the technology stack and possesses specific functionalities, giving rise to various development and deployment challenges.
First, researchers face the complex task of developing a diverse system encompassing various components, ranging from mobile and wearable applications to web servers, requiring diverse programming skills and knowledge.
Moreover, after development, deploying, and maintaining these applications can pose substantial obstacles due to the high frequency, longitudinal nature, and potential scalability of health data streams.
These challenges can impede research progress and divert focus from the core experiments.

Several open-source software platforms have been developed to facilitate mobile health (mHealth) studies \citep{mcerebrum, Radar-base,Bridge}. These platforms offer a range of tools encompassing servers, mobile applications, and analytics tools, providing researchers with diverse possibilities. Researchers can also reprogram these platforms to suit their specific requirements. While mHealth platforms can reduce the need for extensive development, the burden of deployment still rests on the researchers. Additionally, the costs associated with deployment are typically borne by a single organization, making it relatively more expensive for smaller organizations conducting smaller-scale studies.

Alternatively, researchers can utilize online services for conducting their mHealth studies \citep{ilumivu,ethica}. These services are platform solutions provided and deployed by service providers. They are designed to share resources between different organizations and studies, resulting in reduced time and effort required for developing and deploying a custom mHealth application. By sharing resources, these services effectively cut down costs. Typically, these services offer researchers a dashboard for reconfiguring the services with various options. However, the available configurations may not provide the necessary flexibility required for real-time studies.

Despite the significant advancements in existing mHealth solutions, there persists a pressing demand for a comprehensive solution that integrates three essential features into a unified package. Firstly, such a solution should offer a ready-to-use setup that eliminates the requirement for computer programming or infrastructure skills, ensuring accessibility for researchers without technical expertise. Secondly, the solution should prioritize affordability by reducing deployment costs through resource sharing and providing reusable components. Lastly, the solution must exhibit flexibility by offering components that can be combined in various ways to accommodate the diverse and evolving demands of modern mHealth studies, such as personalization.

This paper presents ZotCare, an innovative mHealth service provider that aims to overcome the limitations observed in existing mHealth solutions. We begin by highlighting how ZotCare offers a unique combination of flexibility and programmability, catering to users with diverse skill levels, in contrast to other state-of-the-art solutions. Subsequently, we delve into the details of ZotCare's service orchestration, elucidating how researchers and participants interact with our services to achieve desired outcomes. We specifically explore the capabilities of each service category, including Collection Services, Profile Services, and Real-time Processing, Intervention, and Integration Services, emphasizing the extent of customization achievable by utilizing these services in tandem. Furthermore, we discuss the frontend mobile application of ZotCare, elucidating its role in participant engagement and the provision of personalized experiences within mHealth studies. Additionally, we present the various features of ZotCare's researcher dashboard, showcasing its effectiveness in facilitating study management, service configuration, recruitment, and data analysis. To substantiate the capabilities of ZotCare, we provide multiple use cases and examples of successful mHealth projects that have leveraged ZotCare.

\section{Related Work}

The adoption of mHealth solutions within healthcare applications has witnessed a significant surge, fueled by the shared objective of improving healthcare delivery and outcomes, as highlighted by \cite{McKinsey}. These solutions encompass a wide array of features that greatly facilitate the implementation of mHealth studies. One key aspect is the ability of mHealth solutions to seamlessly integrate wearable devices and diverse data sources, thereby enabling real-time health monitoring. These solutions can also provide data visualization and analytic methods, promote interoperability, and support interventions while ensuring the privacy and security of users.

In the implementation of mHealth solutions, it is crucial to consider and explore three key aspects. The first aspect pertains to the \textit{setup time} required to initiate and configure mHealth studies using the chosen solution. This setup time encompasses various phases, including \textit{system design}, \textit{development}, and \textit{deployment}, each demanding a significant amount of time and effort. These stages involve designing the system architecture, developing the necessary functionalities, and deploying the infrastructure to support the intended mHealth studies.

The second aspect to be considered is the associated \textit{costs} involved in the development and deployment of the mHealth solution. Development costs encompass the investment of human resources and time required for designing and developing the system infrastructure. This includes the efforts of software engineers, data scientists, and other relevant professionals. In addition, ongoing modifications and enhancements may require additional development efforts. Deployment costs encompass the procurement of necessary processing resources, such as servers or cloud infrastructure, as well as ongoing maintenance and operational expenses.

The third aspect revolves around the \textit{customization} capabilities offered by the mHealth solution. Customization can be viewed across three distinct levels: \textit{development}, \textit{configurability}, and \textit{programmability}. At the development level, customization refers to the ability to tailor the solution to meet specific research requirements and objectives. This may involve creating new functionalities or modifying existing ones. Configurability, on the other hand, allows users to adapt the solution's settings and parameters to align with the unique needs of their mHealth studies. Programmability refers to the capability of leveraging programming interfaces or APIs to integrate the solution with other systems or to extend its functionalities.

At the \textit{development} level of customization, researchers are advised to allocate additional efforts to introduce new functionalities and tailor existing mHealth solutions to their specific needs. This level of customization entails direct involvement with the underlying codebase of the solution, thereby necessitating a high level of programming expertise. Researchers must possess the technical skills required to modify the existing code, introduce new functionalities, or make changes to the underlying algorithms.
Moving to the \textit{configurability} level of customization, researchers can customize the solution by reconfiguring the available features within the provided framework. This level of customization does not demand extensive technical expertise and programming skills. Instead, researchers can make adjustments to the system's settings, parameters, or options offered by the solution. While configurability provides a certain degree of customization, it may be limited to predefined configurations and settings, constraining researchers from making substantial modifications beyond the available options. 
Finally, at the \textit{programmability} level of customization, researchers can leverage the solution's programming interfaces or APIs to customize its behavior based on specific situations and conditions. 
In contrast to development-level customization, programmability-level customization offers researchers the ability to incorporate their own functionalities into the system with minimal effort, without requiring extensive technical expertise. In the following, we will provide an overview of existing mHealth solutions, highlight their limitations, and subsequently present the advantages of our solution, ZotCare, in addressing and bridging these gaps. 

\subsection{Related mHealth Solutions}
The existing landscape of mHealth solutions can be broadly categorized into two primary classifications: \textit{platforms} and \textit{services}. \textit{Platforms} encompass comprehensive frameworks that integrate various components of mHealth solutions through the utilization of one or multiple open-source software.
One such platform is Radar-base by \cite{Radar-base}, which focuses on remote monitoring and data collection. It facilitates the integration of data from multiple sensors and devices, enabling comprehensive monitoring capabilities. Another notable open-source platform is mCerebrum by \cite{mcerebrum}, which provides tools for real-time monitoring, data processing, and personalized health interventions based on mobile sensor data. These platforms offer a wide range of features, including data integration, real-time monitoring, analytics, and decision support tools. The Bridge Platform \citep{Bridge} by Sage Bionetworks is another noteworthy example, providing an open-source software framework for digital health research studies. It allows researchers to develop mobile apps, securely collect participant data, and foster participant engagement while emphasizing privacy and data sharing.

However, deploying and utilizing these platforms for mHealth studies require substantial effort, as setting up the necessary software can extend the setup time of studies. Technical challenges may arise, particularly for researchers lacking expertise in Internet infrastructure. Moreover, these platforms are typically designed to operate within a single organization or study, making the deployment costs exclusive to that particular organization. Consequently, this exclusivity can disproportionately affect smaller-scale studies, potentially rendering the deployment financially burdensome.
Another significant challenge associated with these platforms is the limited availability of customization methods. While the open-source nature of these platforms provides some level of customizability at the development level, implementing additional features and functionalities typically necessitates the involvement of technically skilled developers. This dependency on technical expertise may hinder researchers' ability to efficiently add or modify elements within the platform to suit their specific requirements.

Conversely, \textit{services} encompass pre-built solutions that are tailored to specific healthcare needs. These solutions are designed to address particular aspects of healthcare and offer a more focused approach. For instance, ilumivu \cite{ilumivu} provides a closed-source service that facilitates remote patient monitoring and data collection through user-friendly mobile applications. This service emphasizes patient engagement and includes features for symptom tracking, medication adherence, and communication with clinicians. Ethica \cite{ethica}, another closed-source service, places emphasis on privacy-preserving data collection and analysis. It ensures compliance with privacy regulations while enabling remote monitoring and research data collection.

These services offer ready-to-use features and intuitive interfaces, enabling researchers to swiftly adopt and utilize these mHealth solutions without requiring extensive technical expertise. By providing a streamlined and straightforward setup process, these services enable researchers to initiate their studies promptly, leveraging the available features and minimizing setup time.
In terms of costs, services typically entail lower expenses compared to platforms. This is primarily due to the fact that the deployment, maintenance, and resource management burdens are assumed by the service providers. Consequently, these costs are distributed among different studies that utilize the shared resources, making it more cost-effective for researchers.
However, services generally offer limited customizability options, particularly with regard to advanced functionalities. Customization opportunities mainly revolve around configuring existing features to align with researchers' needs. Researchers may encounter limitations when attempting to tailor these services to their specific workflows or integrate additional features beyond those provided by the service.

The choice between platforms and services depends on various factors, including specific requirements, available resources, researchers' technical expertise, and study objectives. Different solutions offer distinct trade-offs in terms of setup time, costs, and customization capabilities.
Services generally offer shorter setup times and lower costs compared to platforms. The pre-built nature of services allows for swift deployment and immediate utilization of the provided features. However, researchers may face limitations in customizing these services to align precisely with their experimental needs. The available configurations may be restricted to the options provided by the service, potentially constraining researchers in their experimentation.
On the other hand, platforms provide a more comprehensive range of customization options. This level of customization, however, typically necessitates expertise in modifying the underlying codebase. 
Furthermore, there is a distinction in the burden and costs associated with deployment and maintenance between platforms and services. With platforms, the responsibility of deployment and maintenance lies with the researchers, entailing additional efforts and costs. In contrast, services assume these burdens on behalf of the researchers, sharing the costs across different organizations utilizing the service. 

\begin{figure*}[ht]
\begin{center}
\includegraphics[width=\textwidth]{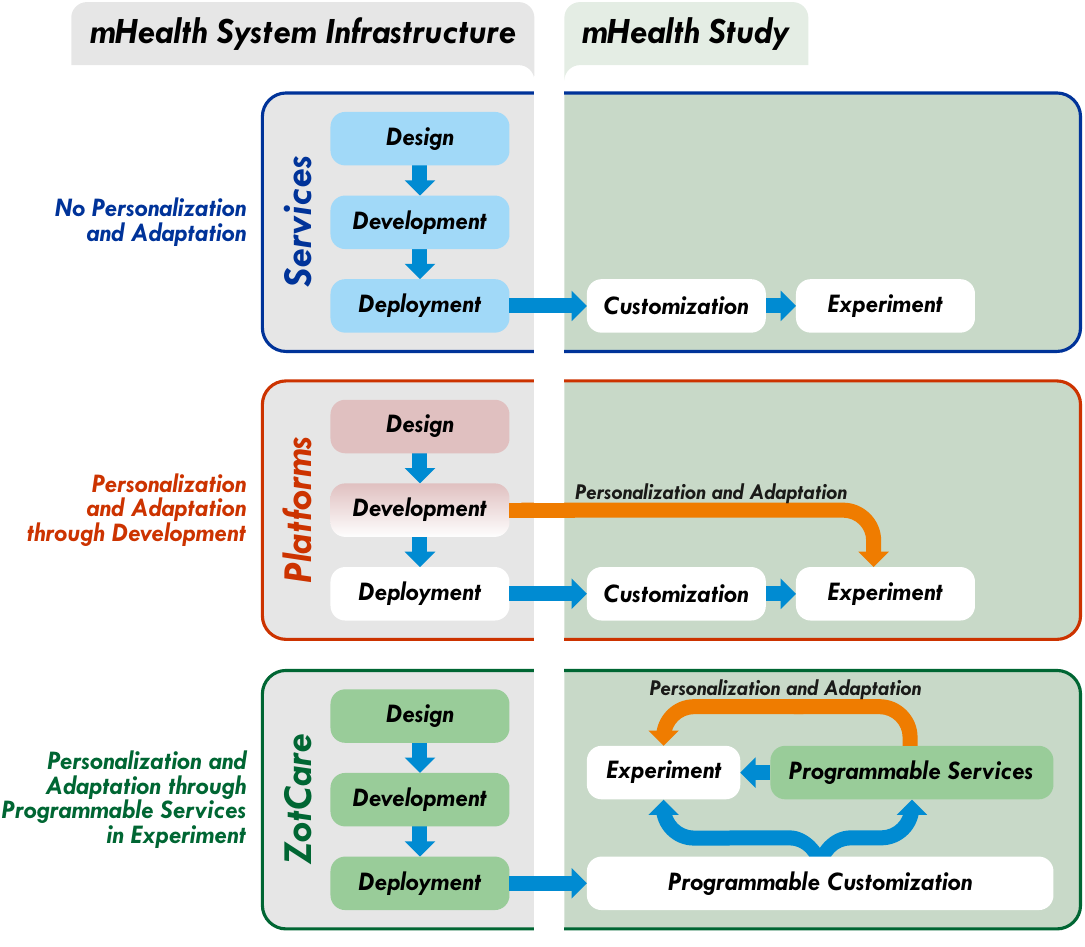}
\end{center}
\caption{mHealth solutions comparison}
\label{fig:comparison}
\end{figure*}

Figure \ref{fig:comparison} provides a comprehensive overview of the key steps involved in conducting a mHealth study and illustrates how platforms and services can aid researchers in each stage of the process. Notably, the figure highlights the advantage of services in facilitating the deployment phase, specifically in building the mHealth system infrastructure. Conversely, services may have limitations in terms of personalization and adaptability, which are addressed by platforms during the system development stage.
Table \ref{tab:comparison} presents a summary of the distinctions between state-of-the-art mHealth solutions, focusing on three key aspects: customization, cost, and setup time. 

Our primary objective is to introduce ZotCare, a comprehensive programmable service orchestration, that combines the advantages of both platforms and services while remaining within the services category. ZotCare is specifically designed to operate within a shared environment, accommodating multiple organizations, studies, and researchers. This shared environment facilitates reduced setup time and costs compared to traditional mHealth platforms.
ZotCare offers extensive customization options across various levels, including development, configurability, and programmability. These customization capabilities allow for seamless implementation of new features and functionalities tailored to specific research needs. Notably, ZotCare excels at the programmability level, providing researchers with a diverse set of tools to achieve the personalization and adaptation required in modern mHealth studies. Figure \ref{fig:comparison} illustrates how researchers can leverage ZotCare's programmable services to attain personalized and adaptive features within their experiments, eliminating the need for additional development efforts.
At the development level, researchers can utilize the open-source version of ZotCare, similar to existing platforms, enabling independent deployment and utilization. 
Table \ref{tab:comparison} summarizes the distinctions between ZotCare and other commonly used platforms and services in the field of mHealth studies. In the subsequent section, we will delve into a detailed discussion of ZotCare's capabilities.

\begin{table}[h]
\centering
\caption{mHealth solutions summary}
\label{tab:comparison}
\resizebox{\textwidth}{!}{
\begin{tabular}{ l | l | l |l | l | l | l }
   & & \multicolumn{3}{c|}{Customization level} &  & \\
  & & Development & Configurability & Programmability & Cost &  Setup time\\ \hline
 
 \multirow{2}{*}{\textbf{Platforms}}& mCerebrum & \checkmark & $\times$ & $\times$ & exclusive & high \\
 
 &Radar-base & \checkmark & $\times$ & $\times$ & exclusive & high \\
 &Bridge & \checkmark & $\times$ & $\times$ & exclusive & high \\
 \hline
 
 \multirow{3}{*}{\textbf{Services}}&ilumivu & $\times$ & \checkmark & $\times$ & shared & low \\
 &Ethica & $\times$ & \checkmark & $\times$ & shared & low \\
 &\textbf{ZotCare} & \checkmark & \checkmark & \checkmark & \textbf{shared} & \textbf{low}  \\
\end{tabular}
}
\end{table}

\section{ZotCare Service Orchestration}

ZotCare constitutes a Health Cybernetics platform, specifically designed to operate as a closed-loop real-time monitoring-intervention system. Its purpose is to cater to the requirements of researchers, clinicians, and community health workers engaged in conducting studies or delivering digital health services. This comprehensive platform enables ubiquitous monitoring of individuals, encompassing both general populations and those at heightened risk, while also providing mHealth interventions. Additionally, it offers a direct avenue for end-users to engage in self-management.

ZotCare encompasses fundamental components essential for conducting mHealth studies across the entire health technology stack. Notably, it provides services that streamline data collection through the utilization of intelligent devices, such as wearables and portable devices. Furthermore, it enables bidirectional interactions between study participants and researchers through gateway devices, including smartphones. Augmenting its capabilities, ZotCare's cloud services provide data analysis and visualization, facilitate the construction and execution of real-time predictive models, and initiate actions necessary to enable just-in-time adaptive interventions (JITAI).

The primary aim of ZotCare is to enable the expeditious and convenient advancement of mHealth solutions catering to users possessing varying degrees of programming and engineering expertise, irrespective of their level of technological literacy. Consequently, through utilizing ZotCare services, researchers can efficiently diminish the time and expenses associated with the implementation and deployment of monitoring systems, enabling them to focus their endeavors on study design, conceptualization, and participant engagement.

\begin{figure*}[ht]
\begin{center}
\includegraphics[width=\textwidth]{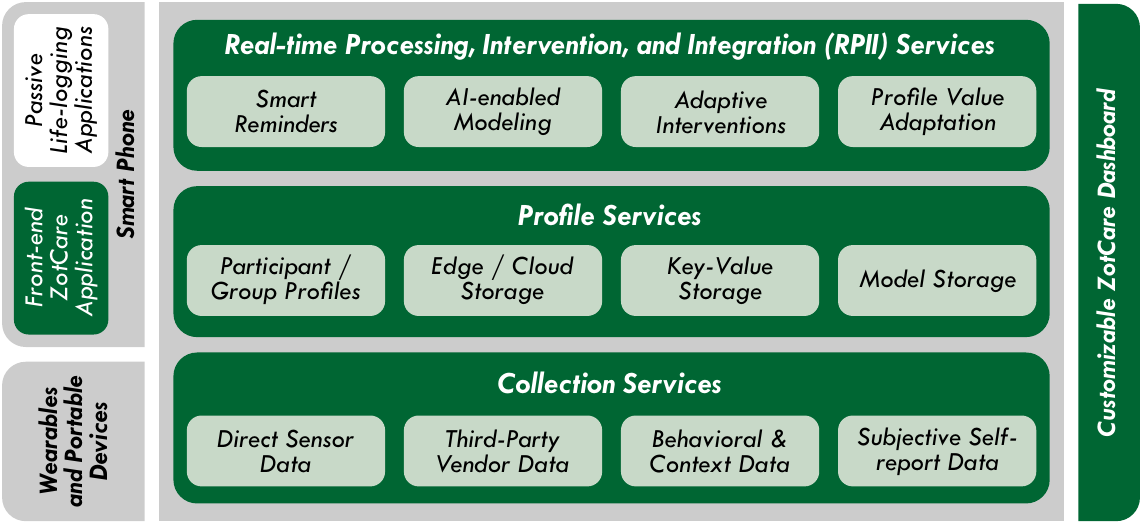}
\end{center}
\caption{ZotCare services overview}
\label{fig:services}
\end{figure*}

Figure \ref{fig:services} illustrates a comprehensive overview of ZotCare services and interfaces.
The \textit{Data Collection Services} facilitate the ingestion of data from diverse devices, applications, and services. Once collected, the data undergoes processing and is stored as a continuous stream within ZotCare.
The \textit{Profile Services} assume responsibility for the storage and processing of data in the form of key-value pairs. This storage mechanism enables the creation of profiles for participants and groups, serving as a repository for personalized study-related data and models, as further elucidated subsequently.
Through the \textit{Real-time Processing, Intervention, and Integration (RPII) Services}, researchers possess the capability to incorporate adaptive, intelligent, and real-time components into their studies. These components are capable of triggering various actions based on the data obtained from the \textit{Profile} and \textit{Collection} services.
In conjunction with these services, ZotCare provides two interfaces: a customizable dashboard and a user-facing mobile application.
The customizable ZotCare dashboard serves as a web application, offering researchers an interface for accessing and modifying ZotCare services pertinent to their respective studies. Researchers can employ the dashboard to manage collected data, recruit participants, and customize it for clinical purposes if desired.
The ZotCare application, on the other hand, functions as a user-facing mobile application for participants. It allows them to interact with ZotCare services, enabling functionalities such as receiving reminders, engaging in ecological momentary assessments (EMAs), and benefiting from adaptive mobile health interventions. Moreover, ZotCare facilitates the integration of contextual and behavioral monitoring applications, commonly referred to as lifelogging applications.
The subsequent subsections delve into further details regarding ZotCare services and provide insights into how researchers can effectively leverage these services to construct their closed-loop mHealth solutions.

\subsection{Collection Services}

The \emph{Collection Services} assume the responsibility of acquiring and integrating participants' data within ZotCare.
Given the multifaceted nature of mHealth studies, various types of data are typically employed.
Objective physiological, behavioral, and contextual data, alongside subjective self-reported data, constitute the principal data types utilized in the context of mHealth studies.
Furthermore, third-party vendors and applications offer diverse methodologies for data collection, encompassing direct sensor readings as well as indirect data acquisition through their server-side APIs.
To accommodate these disparate data types and collection methods, ZotCare incorporates a range of features that enable the acquisition of data through diverse channels, subsequently presenting them to researchers in a cohesive and standardized format.

The Collection Service possesses the capability to gather physiological data from prominent fitness and well-being devices. These devices encompass wearable options, such as smartwatches and rings, and portable devices, like smart blood pressure monitors and scales.
These devices are capable of providing physiological data in processed formats, and in some cases, as raw data. The raw data typically comprises inertial measurements (accelerometer and gyroscope), photoplethysmography (PPG), electrocardiogram (ECG), air pressure, luminosity sensor data, and other sensor readings, contingent upon the specific type and model of the device.
On the other hand, processed data generally entails higher-level derived physiological metrics such as heart rate, heart rate variability, sleep quality, steps, exercise data, weight, and other relevant parameters. These metrics are derived from the raw sensor readings by the respective vendors.

To facilitate the collection of such data, ZotCare has been integrated with various healthcare device vendors. Presently, ZotCare offers support for Samsung, Garmin, Empatica, and Fitbit smartwatches, as well as Oura rings for smart wearables. Additionally, ZotCare can integrate with Withings smart scales and blood pressure monitors.
It is important to note that the list of supported devices is continually expanding, as indicated in Table \ref{tab: HRVMeasures}. For certain devices that provide a software development kit (SDK) and open access to their operating system/firmware (e.g., Samsung Active watches running Tizen OS), ZotCare offers a native smartwatch application. This application enables direct access to the raw signals from these devices and transmits them to the ZotCare back-end.
Researchers also have the flexibility to incorporate new devices through direct connections or by utilizing third-party services, utilizing standard open authentication (OAuth) methods.

Furthermore, to augment data collection capabilities within ZotCare, we have seamlessly integrated the AWARE smartphone-based logging framework \citep{aware-framework} to enable the passive collection of behavioral and contextual data.
Through AWARE, researchers can leverage participants' smartphones to gather data from various sensors, including location, accelerometer, battery status, light intensity, temperature, and more. AWARE also allows the extraction of contextual information from participants' daily lives, such as screen lock/unlock events, application usage patterns, step count, and even communication activities such as notifications, text messages, and phone calls.

To encompass the collection of self-reported subjective data within ZotCare, we have incorporated an \textit{Interaction} sub-service into the system. This feature empowers researchers to design and deploy dynamic questionnaires, indicators, and interactive tasks using the Interaction's functionality. The ZotCare front-end application effectively handles these Interactions, capturing participants' responses along with detailed metadata for comprehensive analytics. Moreover, the Interactions feature serves as a versatile tool for various purposes, including EMAs, information delivery, assessments, recommendations, and interventions. Researchers have the flexibility to update questions, EMAs, and other interactive components on-the-fly using the ZotCare dashboard, granting them dynamic control over the study's data collection processes. 

\begin{table}[h]
\centering
\caption{ZotCare objective data collection}
\label{tab: HRVMeasures}
\begin{tabular}{ l | l | l | l }
\textbf{Device}  & \textbf{Data type}  & \textbf{Integration type} & \textbf{Dev stage} \\ \hline
Samsung Tizen watches          & raw/processed              & direct                & supported                \\
WearOS-enabled watches (Fossil, Pixel, etc.) & raw/processed  & direct                & under-dev          \\
Empatica E4 wristband & raw/processed  & direct/third-party                & under-dev          \\\hline
Garmin           & processed            & third-party     & supported                \\
Whitings devices (BP, Scale, etc.)         & processed            & third-party     & supported                \\
Fitbit           & processed          & third-party     & under-dev                \\
Oura             & processed    & third-party                & supported \\ \hline
AWARE  & raw/processed      & direct                & supported                \\ 
\end{tabular}
\end{table}

\subsection{Profile Services}

The Profile Services within ZotCare assume the responsibility of storing specific information pertaining to groups or individual participants. Researchers can program these profiles to establish key-value storage for data management purposes.
In the case of participant profiles, the programmed key-value storage consists of a predetermined set of keys established by the researchers for all participants. However, individual values can be stored per key for each participant, allowing for personalized data storage.
For group profiles, a single value is associated with each key, which can be replicated across different groups. This replication enables the creation of distinct groups, such as control and intervention groups, or allows for customization of shared resources, such as the ZotCare Frontend application.
Each key within the profiles can be configured with a variety of features. Researchers have the flexibility to choose whether the values associated with these keys should be stored on participants' edge devices or in the cloud. Additionally, researchers can determine whether these values should be visible to the participants, depending on the study's specific requirements and privacy considerations.

The Profile Services play a crucial role in enabling researchers to personalize and adapt their studies over time, particularly in advanced studies that require participant engagement, personalized interactions, or the utilization of statistical or AI models. However, studies that primarily focus on monitoring and passive data collection may not extensively utilize this service.
Within participants' profiles, researchers can store a range of important dates and times, such as join date, delivery date, significant personal events, and preferred notification times. Additionally, characteristics such as height, weight, and fitness level can be recorded. Serializable entities, such as personal AI models or statistical models, as well as files like images, audio recordings, or voice recordings, can also be stored within participants' profiles.

Group profiles, on the other hand, contain information that is shared among the members of a specific group. This may include timing information for different stages of the study or shared AI models. Furthermore, group profiles can include customization data specific to each study, such as differentiating between intervention and control groups or specifying menu items. The information stored within profiles serves multiple purposes within the \textit{Real-time Processing, Intervention, and Integration Services}. It allows for the adaptation of study procedures based on individual participant characteristics. Researchers can also leverage profile information within Interactions to customize and personalize the individual experiences of participants. Furthermore, profiles can be used to locally store personal identifiers such as names, addresses, and photos, instead of saving them on servers. This enables further customization of the participant's experience while preserving their privacy.

Overall, the Profile service provides researchers with a versatile tool for personalization, adaptation, and customization, enhancing the effectiveness and participant-centric nature of their studies.

\subsection{Real-time Processing, Intervention, and Integration (RPII) Services}

ZotCare offers researchers a comprehensive suite of \textit{Real-time Processing, Intervention, and Integration (RPII) Services}, which equip them with the capability to transform data into knowledge, incorporate intelligence into their studies, and effectively close the loop within their solutions.

Through the RPII Services, researchers gain the ability to process data derived from the Profile and Collection services, enabling them to extract meaningful insights and execute subsequent actions based on the processed data. These services can be leveraged at various stages of the data processing pipeline, encompassing tasks such as data pre-processing, AI model development, collection of smart labels and EMAs, scheduling adaptive interventions, and sending intelligent reminders.

By utilizing the RPII Services, researchers are empowered with complete control over the flow of data within their studies. This enables them to dynamically analyze and respond to data in real-time, facilitating the integration of intelligence into their research and ultimately closing the loop within the solution they have developed.

Within ZotCare, each study is capable of containing multiple Real-time Processing, Intervention, and Integration (RPII) instances, which play a pivotal role in enabling dynamic and intelligent functionality. Each RPII instance consists of three essential components: \textbf{\textit{Triggers}}, \textbf{\textit{Conditions}}, and \textbf{\textit{Actions}}.
\textit{Triggers} serve as indicators that determine when an RPII unit is to be executed. These triggers can be categorized as either data-driven, responding to incoming new data, or chronological, based on fixed times or frequencies.
\textit{Conditions}, on the other hand, evaluate the data to determine if any adaptations or actions need to be performed. Based on the specified conditions and the available data, the RPII instance can make informed decisions regarding the subsequent actions.
\textit{Actions} within an RPII instance are programmable functions that can trigger internal modifications within the ZotCare environment or invoke external functionalities. Researchers have the flexibility to program RPII instances with various internal functions within ZotCare, including data fetching, participant grouping and filtering, data processing, AI model building, and writing to data streams or profile values. Furthermore, ZotCare supports external actions such as sending emails, push notifications to the ZotCare mobile application, and accessing external resources.
To provide an overview of these features, a comprehensive summary is presented in Table \ref{tab: logic features}, which outlines the various logic features supported by ZotCare.

\begin{table}[h]
\centering
\caption{Zotcare RPII services features}
\label{tab: logic features}
\begin{tabular}{l|l|l|l}
\textbf{Component}  & \textbf{Type} & \textbf{Options} & \textbf{Dev stage} \\ \hline
Triggers & Data & incoming data & supported \\
& Chronological      & cron expressions               & supported          \\ \hline
Conditions & Fetch             & data streams \& profiles                   & supported          \\
           & Filter            & data streams \& profiles            & supported          \\
           & If / Else         & -                 & supported          \\
           & Inferring AI models      & -                 & under-dev \\
           \hline
Actions    & Send Email        & templates \& plain & supported          \\
           & Send Push Notification & 
           \begin{tabular}{l}
           ZotCare \&\\ Firebase \cite{firebase} \&\\ OneSignal \cite{onesignal} \end{tabular} & supported          \\
           & Write Profile & - & supported  \\
           & Training AI models      & -                 & under-dev 
\end{tabular}
\end{table}

Moreover, ZotCare offers seamless integration options for external systems with its RPII services. Researchers are provided with dedicated endpoints to access ZotCare from their own machines and servers, facilitating the integration of external resources into the ZotCare environment. To streamline the process of utilizing ZotCare externally, an SDK is available, designed to simplify the interaction with ZotCare and offer additional features. The SDK enables researchers to fetch, cache, and process data, as well as invoke actions within ZotCare, all without the need to handle complex authentications or intricate API calls.
By leveraging these integration capabilities, researchers can utilize their own resources to replace or supplement ZotCare's RPII components, enhancing the flexibility and adaptability of the system to suit their specific requirements. 

\subsection{The Customizable ZotCare Dashboard}

The ZotCare dashboard serves as a customizable interface that facilitates interaction between users (such as researchers and clinicians) and ZotCare services.
Researchers can create different study groups through the dashboard. 
Each group can be configured to utilize the ZotCare services for the purpose of that specific research or product.

The ZotCare dashboard incorporates a dedicated section for user management. Within this section, researchers can recruit new participants for their studies. This can be achieved through the utilization of random IDs for direct recruitment or by utilizing sign-up links for anonymous recruitment. Additionally, the dashboard enables researchers to edit user information and profile values as needed.
Furthermore, the ZotCare dashboard offers a comprehensive suite of data analysis capabilities, ensuring that researchers have the necessary tools to derive valuable insights from their research data. Researchers can leverage the provided tools to visualize data in its original format or apply sophisticated aggregation and filtering techniques to create visually informative charts and graphs. Moreover, the dashboard empowers researchers to employ their domain knowledge and expertise by facilitating direct annotation of data within the platform. These annotations are seamlessly stored as new data streams within ZotCare, contributing to a rich and comprehensive dataset for further analysis.

In addition to user management and data analysis functionalities, the ZotCare dashboard provides researchers with effective tools for managing their services within the platform. Through an intuitive interface, researchers can easily activate, modify, or review the configurations of their services. While certain services, such as collection services, entail straightforward setup steps, others, such as programmable services like RPII, profile, and interactions services, necessitate more advanced configurations. To streamline this process, the dashboard offers interactive editors that facilitate researchers in editing, debugging, and testing these programmable services, ensuring a seamless and efficient management experience.

ZotCare also incorporates a fine-grained access control mechanism that allows users to have specific permissions within individual studies. This feature enables researchers to involve different collaborators in their study, assigning them distinct roles based on their access scope. These roles can range from recruiters or data analysts to study managers or clinicians. Each collaborator is granted access only to the relevant parts of the dashboard that align with their assigned role. This stringent access control is crucial for safeguarding the privacy and integrity of the study, ensuring that each collaborator can only view and utilize the components that pertain to their specific responsibilities.

\subsection{ZotCare Mobile Application}

The ZotCare mobile application serves as an interface for facilitating ZotCare services to participants. This mobile app functions as a front-end interface, enabling various services, including mHealth interventions, multimedia interactions, and interactive profiles through its components.
Additionally, the ZotCare app acts as an assistant to participants, aiding them in device setup and facilitating communication between participants and researchers/clinicians.
The primary purpose of the app is to provide participants with interactive "interactions." These interactions encompass a range of components, such as multiple-choice, numerical, time, data, and text input, as well as sliders, among others. These components are well-suited for various purposes, such as EMAs, questionnaires, and data labeling, which are commonly employed in mHealth studies.
Furthermore, interactions are equipped with multimedia features, including videos, images, audio, and audio-video recorders. Extensive research has demonstrated the effectiveness of these multimedia tools for both assessment and mHealth interventions, as evidenced by studies presented in Section \ref{sec:use-cases}.

Moreover, interactions can incorporate customized components that researchers can create and incorporate, allowing for further customization and enhancement of their studies. Previous studies using ZotCare have showcased the utilization of such components for interventions, such as interactive breathing exercises, mindfulness-oriented image galleries, relational savoring exercises, and educational materials. Additionally, these components have been employed in assessments, such as cognitive games (e.g., finger tapping, word pair memory tests, rule-switching games, etc.).
In addition to the visible components, interactions can include condition statements, representation configurations, variables, and metadata. These features provide researchers with a broader set of tools for personalization and customization.

Furthermore, participants have the ability to grant authorization to ZotCare, via the application, to access their health data from third-party services, such as Oura and Garmin. This integration allows for seamless retrieval of pertinent health information. 
Additionally, the app offers comprehensive instructions and troubleshooting steps for devices and applications that establish a direct connection with ZotCare, including Samsung and AWARE.
Furthermore, participants can access certain features of the Collection and Profile services through the ZotCare app. These services provide participants with valuable functionalities and data management capabilities.
The ZotCare app serves as a means for researchers and participants to maintain a continuous connection. This connection is facilitated through various means, including reminders, notifications, and messages. Researchers can choose to automate these communications through the RPII services or manually trigger them using the dashboard. 

A general version of the ZotCare app is readily available for installation and use on Android and iOS smartphones. However, the app's flexibility allows for customization to accommodate different research studies. Researchers possess the capability to modify the app's colors, logos, menus, and other visual aspects to align with the specific requirements of their study. Moreover, they can also modify the app's components to create tailored "Interactions" with additional functionalities. By leveraging the Profiles feature, researchers can further personalize the app's appearance to suit individual studies or specific participant requirements.

\subsection{Security and Privacy}

In order to uphold the integrity of ZotCare's services, it is imperative to prioritize the security and privacy aspects of the platform. Robust security measures have been implemented to safeguard data and communication channels against potential threats posed by unauthorized individuals attempting to manipulate, delete, or disrupt data storage and transmission processes.
Privacy considerations within ZotCare are designed to empower participants by granting them control over their personal data. This includes ensuring the protection of identifiable information, thereby safeguarding the privacy of participants.
Security and privacy present unique challenges within service-based environments compared to platforms, as multiple organizations share the same resources. Consequently, it becomes necessary to implement measures to safeguard information from both internal and external sources.

Collected data from participants may consist of both objective sensitive data, such as location information and passwords, as well as subjective data that, based on responses provided in Interactions and Profiles, may reveal sensitive information. Similar security and privacy risks exist across other services as well. For instance, programmable services, including RPII services, are susceptible to malware injection. It is worth noting that researchers' mistakes, such as data overwriting, large or repeated queries, and infinite loops, can also introduce malware vulnerabilities.

To address these security challenges, ZotCare has implemented a gateway service that regulates authentication, authorization, and scope through standard encryption methods. This entails a two-step process in which the gateway first verifies the identity of the requester and subsequently checks if the requester has the necessary access permissions to the requested resource.
Privacy concerns extend beyond the scope of data collection and storage and begin with participant recruitment. In cases where studies possess knowledge of their participants, researchers can manage deidentification processes on their end, enrolling participants in ZotCare using anonymous IDs. However, for studies that allow individual participant sign-ups, ZotCare can deidentify data associated with participant emails, enabling participants to utilize their emails for password retrieval and receiving notifications. Nevertheless, researchers only have access to anonymous IDs.
ZotCare does not currently support deidentification of collected data at this stage. It is important to note that both researchers and users have the option to disable the collection of sensitive data across all ZotCare services, providing an additional layer of privacy control.

\section{Use cases} \label{sec:use-cases}

ZotCare has been utilized as a service within diverse mHealth research studies. 
The functionalities of ZotCare were devised to meet the requisites of these studies and were adapted accordingly based on their specific utilization. 
The initial studies availed themselves of preliminary versions of ZotCare, encompassing provisions for multi-modal data collection. 
Subsequently, ZotCare broadened its spectrum of services and characteristics to address the requirements for customization and governance. 
In the following, we will begin by providing an overview of select studies that used ZotCare services for purposes encompassing data collection, data modeling, and intervention. 
Subsequently, we will delineate the challenges encountered and describe the integration of ZotCare into these aforementioned studies.

\subsection{Personal Mental Health Navigation Project}

The Mental Health Navigation (MHN) project develops a proactive, personalized approach to monitor, estimate, and guide individuals toward their desirable mental health state \citep{mhn_paper}.
MHN monitors a multimodal stream of objective and subjective information to build inference models to determine participants' mental states, context, and lifestyle.
Using the constructed personal model and the current state, a navigator system can steer the participants using interventions at each step. 
The MHN project comprised two studies, the \emph{Affect} study and the \emph{Loneliness} study.
The \textit{Affect} study focused on investigating the connection between college students' psychophysiological signals and sleep on their mood \citep{mhn-affect-technical-paper}.
Due to the onset of the COVID-19 pandemic during the midst of the \textit{Affect} study, a revision was made to expand the study's objectives to encompass the impact of COVID-19 and subsequent lockdown measures on the lives and emotional well-being of college students \citep{mhn-sleep,mhn-context}.
The subsequent phase of the MHN project referred to as the \textit{Loneliness} study, was primarily dedicated to the real-time evaluation of the mental well-being of college students, along with the provision of just-in-time adaptive interventions for those individuals requiring support \citep{mhn-loneliness}. 
Moreover, the \textit{Loneliness} study encompassed the collection of life-logging and contextual data, enabling the inference of participants' virtual (through smartphones) and physical communication levels. 
By integrating the acquired life-logging and contextual data with the pre-existing models established in the \textit{Affect} study, the accuracy of the loneliness assessment models was significantly enhanced. 
Consequently, the Loneliness study successfully developed adaptive interventions, leveraging these refined models, with the ultimate aim of mitigating the adverse mental state experienced by the participants.

Both the \textit{Affect} and \textit{Loneliness} studies employed the utilization of ZotCare as a means to gather bio-signal and EMA data and administer mHealth interventions. 
The \textit{Affect} study specifically utilized the Samsung smartwatch and Oura ring to continuously capture physiological signals, monitor sleep quality, and track levels of physical activity. 
In addition to these devices, the \textit{Loneliness} study incorporated the use of AWARE to gather life-logging data. 
Furthermore, a customized ZotCare mobile application, namely mSavorUs, was developed and employed for the participants. 
This tailored application not only retained all the features of the original ZotCare application but also provided supplementary custom features for relational savoring exercises and interventions.
The interventions within the mSavorUs application were personalized by incorporating participants' names and photos sourced from their memories, utilizing locally stored profiles. 
To implement the interventions in the \textit{Loneliness} study, the RPII services were utilized. 
These services were employed both directly for delivering conditional notifications and via an API form to establish a connection with a machine learning agent maintained within a separate cluster.
The volume of data collected in each of these studies exceeded 200 gigabytes, with a total of 4.5K and 14.9K labels collected for the \textit{Affect} and \textit{Loneliness} studies, respectively.

\begin{table}[ht]
\centering
\RaggedRight
\caption{MHN project studies summary}
\label{tab: mhn-studies}
\begin{tabular}{p{.09\textwidth}|l|p{.08\textwidth}|p{.15\textwidth}|p{.15\textwidth}|p{.30\textwidth}}

\textbf{Study} & \textbf{\#} & \textbf{Duration} & \textbf{Collection Service} & \textbf{Profile Service} & \textbf{RPII Service} \\ \hline

MHN Affect \citep{mhn-affect-study} & 20 & 3-12 months & 
$\bullet$ Samsung watch \newline $\bullet$ Oura ring \newline $\bullet$ ZotCare app & 
- & -  \\ \hline

MHN Loneliness \citep{mhn-loneliness} & 20 & 1 year & 
$\bullet$ Samsung watch \newline $\bullet$ Oura ring \newline $\bullet$ ZotCare app \newline $\bullet$ AWARE & 

Local (name and memory images)
& $\bullet$ Build mood prediction models \newline $\bullet$ Trigger interventions based on mood \newline $\bullet$ Send conditional reminders  \\

\end{tabular}
\end{table}

The initial deployment of ZotCare for the MHN \textit{Affect} study presented various challenges that underscored the necessity for additional features or services to address them effectively. 
Given the study's incorporation of multiple data collection dimensions, including the Samsung smartwatch, Oura ring, AWARE framework, and questionnaires, study coordinators faced difficulties in monitoring the status of all dimensions and promptly informing participants about any potential issues and interruptions in data collections.
Interruptions occasionally occurred in the ZotCare collection services running on participants' devices due to factors such as high battery consumption, device inactivity, or inadvertent shutdowns by participants. 
To overcome this challenge, ZotCare implemented a summary report generator within the data collection service. 
These daily summaries could be utilized by researchers through ZotCare dashboards to assess the status of participant data and facilitate timely follow-up when necessary.
Additionally, interactive troubleshooting components were integrated into ZotCare Interactions, accessible via the ZotCare mobile application. 
These troubleshooting components systematically analyzed the participants' collected data and provided step-by-step instructions for resolving any data communication issues that may have arisen.
Through the implementation of these additional features and services, ZotCare effectively tackled the challenges encountered during the MHN \textit{Affect} study. 
This ensured efficient data monitoring, facilitated effective troubleshooting, and ultimately enhanced participant engagement and study outcomes.
Moreover, during the course of the \textit{Loneliness} study, the construction of the real-time inference model necessitated computational resources that surpassed the capacity of the ZotCare backend services available at that time. To address this challenge, the researchers leveraged the RPII capability of ZotCare, enabling integration with external resources. The team successfully employed the RPII service to execute an inference model within a cluster, which was triggered by ZotCare web-hooks. The obtained data was subsequently retrieved through the ZotCare SDK, processed, and intervention scheduling was performed using ZotCare API calls. This strategic approach effectively resolved the issue of resource limitations and facilitated the seamless integration of just-in-time adaptive interventions by harnessing the capabilities of external resources.

\subsection{The UNITE Project}

Smart, Connected, and Coordinated Maternal Care for Underserved Communities (UNITE) is a research project funded by the United States National Science Foundation, with the primary objective of developing innovative technologies to enhance the physical and emotional well-being of underserved pregnant women and their newborns.
The UNITE initiative endeavors to revolutionize conventional maternal care practices, which have traditionally been delivered within homes or clinics, by integrating an AI-supported remote monitoring system. The project comprises three distinct phases, each with its specific focus and objectives.
The initial phase, known as the ``Feasibility" phase, concentrated on assessing the viability of remote maternal health monitoring. This involved investigating the level of engagement exhibited by pregnant women with the technology, taking into account their individual health conditions \citep{unite-feasibility}.
The second phase of UNITE encompassed a series of small-scale randomized controlled trials (MicroRCTs), which sought to examine various aspects of maternal well-being, such as stress management or physical training, and assess their impact on pregnancy outcomes \citep{tazarv-study}.
Currently, the project is in its third phase, an "AI-assisted" study aimed at exploring the efficacy of incorporating AI assistants and nurses into the care loop for mothers. Within this phase, an AI-enabled exercise recommendation system has been deployed alongside the recommendations provided by healthcare providers. This human-in-the-loop mHealth approach has resulted in the development of a personalized, step-by-step recommender system that adapts to the specific pregnancy conditions and physical measures of each individual mother.

\begin{table}[h]
\RaggedRight
\caption{UNITE project studies summary}
\label{tab:unite-studies}
\centering
\begin{tabular}{p{.14\textwidth}|p{.1\textwidth}|p{.1\textwidth}|p{.17\textwidth}|p{.17\textwidth}|p{.17\textwidth}}
\textbf{Study} & \textbf{\#} & \textbf{Duration} & \textbf{Collection Services} & \textbf{Profile Services} & \textbf{RPII Services} \\ \hline

UNITE Feasibility Study \citep{unite-feasibility} & 25 & 6-10 months & $\bullet$ Samsung watch \newline $\bullet$ Oura ring \newline $\bullet$ ZotCare app & - & $\bullet$ Triage system for early risk assessment \\ \hline

UNITE Stress Detection \citep{tazarv-study} & 14 & 30 days & $\bullet$ Samsung watch \newline $\bullet$ ZotCare app & - & $\bullet$ Using stat models to trigger EMAs \\ \hline

UNITE Stress Detection \citep{tazarv-active} & 18 & 2 months & $\bullet$ Samsung watch  \newline $\bullet$ ZotCare app \newline $\bullet$ AWARE & - & $\bullet$ Using reinforcement learning to trigger EMAs \\ \hline

UNITE Exercise Recommender System & 20 & 4 months & $\bullet$ Samsung watch \newline $\bullet$ Oura ring \newline $\bullet$ ZotCare app & $\bullet$ Create a physical profile (height, weight) \newline $\bullet$ Indicators (exercise, set, repetition, duration, intensity) \newline $\bullet$ AI suggestions \newline $\bullet$ Store models & $\bullet$ Triage system for risky behaviors \newline $\bullet$ Exercise recommender engine
\end{tabular}
\end{table}

The Feasibility phase of the UNITE project primarily focused on the collection of data and subsequent analysis during the post-study phase. Given the vulnerable nature of the study group consisting of pregnant mothers, it was imperative to implement a triage system to swiftly identify and report any potentially risky situations. To fulfill this requirement, the UNITE initiative incorporated ZotCare's RPII services. By integrating data-driven triggers within the RPII unit, the behavior of participants could be assessed, enabling immediate alerts to be sent to researchers when necessary.

In the second phase of UNITE, known as the MicroRCT phase, the team explored an alternative approach to collecting labels. Instead of adhering to fixed times and frequencies, the possibility of employing smart labels was investigated, aiming to maximize the information obtained while minimizing participant interruptions. By utilizing statistical and active machine learning models within the RPII services, the UNITE team could send notifications to participants, prompting them to provide labels at more opportune times, resulting in improved accuracy and heightened participant engagement \citep{tazarv-active}.

\begin{figure*}[ht]
\begin{center}
\includegraphics[width=\textwidth]{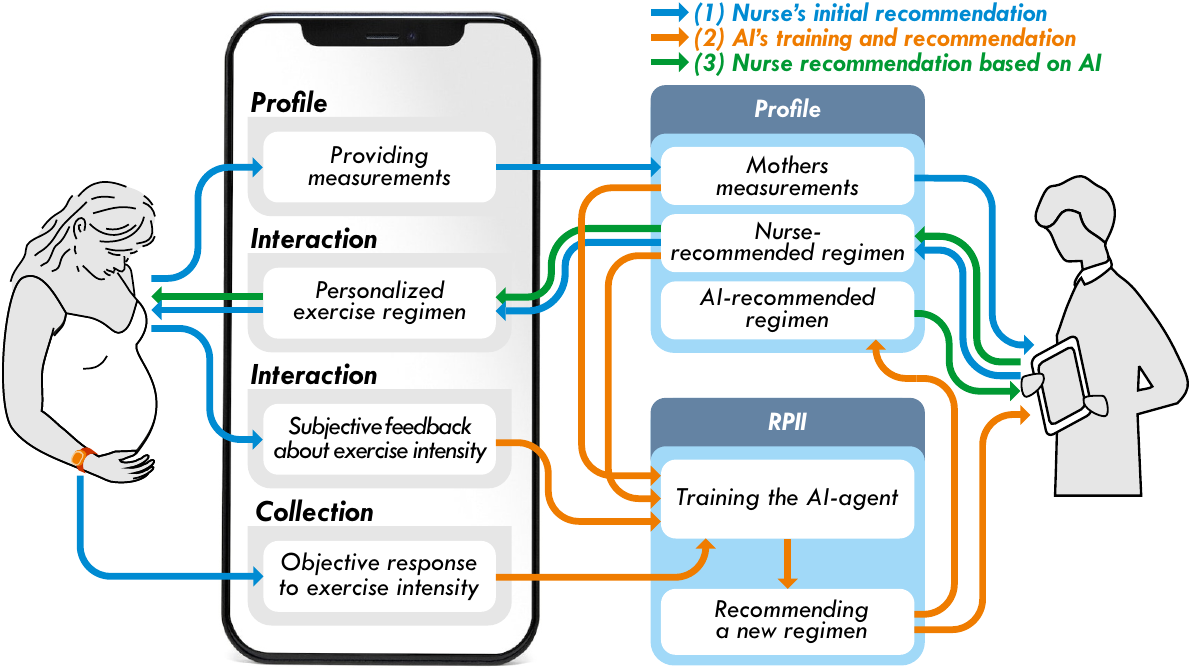}
\end{center}
\caption{UNITE AI-assistant integration in ZotCare}
\label{fig:unite-ai-assisstant}
\end{figure*}

During the AI-assisted recommender system phase, the main challenge revolved around establishing a cohesive loop involving the participants, the AI recommender engine, and health providers. To address this challenge, the Profile and RPII services were leveraged, providing the necessary flexibility.
Three sets of profiles were employed: one for participants to input their physical measurements, another for health providers to input their assessments, and a final one for the AI recommender engine to store its recommendations. 
The RPII service played a crucial role in executing the recommender engine by utilizing the physical measurements and health providers' assessments from the Profile services, as well as participants' bio-signals, progress, and feedback from the Collection services.
Figure \ref{fig:unite-ai-assisstant} illustrates the adoption of the services within this solution. The process begins with Step (1), where participants input their physical measurements through the mobile app. This information is then used by the nurses to recommend the participants' initial exercise regimen.
In Step (2), the recommender engine utilized this information to train its models, infer the next set of exercise regimen recommendations, store the recommendations within the participant's profile, and notify the health providers of the ongoing process. The health providers, in Step (3), evaluated the final exercise regimen for each participant based on the suggestions provided by the recommender engine.
Steps (2) and (3) continue in a continuous loop throughout the duration of the study.
This system effectively assists health providers in processing a significant amount of information and providing frequent assessments.

\subsection{Other Projects}

As the services provided by ZotCare continued to develop, they began to be utilized by researchers and universities outside of the original project, encompassing a diverse array of research studies. These studies varied in terms of their contextual settings, languages used, time zones, and specific requirements. The expanding scope of applications enabled ZotCare to adapt and enhance its functionalities, thereby providing researchers with a wider range of features to facilitate their studies effectively. Some example studies are listed in Table \ref{tab:studies}.

\begin{table}[h]
\RaggedRight
\caption{Examples of other studies that utilized ZotCare}
\label{tab:studies}
\resizebox{\textwidth}{!}{%
\begin{tabular}{p{.12\textwidth}|p{.25\textwidth}|l|p{.09\textwidth}|p{.15\textwidth}|p{.11\textwidth}|p{.15\textwidth}}
\textbf{Study} & \textbf{Description} & \textbf{\#} & \textbf{Duration} & \textbf{Collection Service} & \textbf{Profile Service} & \textbf{RPII Service} \\ \hline

Sleep \& Menstration \citep{negin-study} & Subjective study on the affect of menstruation on sleep & 20 & 2 months & ZotCare app (customized with cognitive tests) & - & - \\ \hline

PREVENT \citep{finland-slim-study}          & Daily well-being of pregnant women around COVID-19 pandemic & 38 & 30 days & $\bullet$ Samsung watch \newline $\bullet$ ZotCare app (in Finnish) & - & - \\ \hline

D-CCC \citep{dccc-paper} & D-CCC was proposed to assist community organiztions to monitor elderly & 5 & 2 months & $\bullet$ ZotCare app\newline $\bullet$ Oura ring \newline $\bullet$ Withings Blood Pressure Monitor \newline $\bullet$ Withings Scale &     -    &  $\bullet$ Interventions to promote physical and mental well-being \\ \hline

Sleep \& Brain study & Impact of sleep quality on memory and cognitive performance & TBD & TBD & $\bullet$ ZotCare app (customized with cognitive tests) & $\bullet$ Custom notification times & $\bullet$ Notification scheduler 

\end{tabular}}
\end{table}

Two research projects, namely ``Sleep \& Menstruation" \citep{negin-study} and ``Sleep \& Brain," were conducted to investigate subjective sleep assessment and cognitive abilities through the use of questionnaires and cognitive tasks. 
The first project examined the impact of menstruation on sleep patterns, while the second project explored the relationship between sleep and cognitive abilities.
Both projects incorporated traditional questionnaires to gather information on various aspects of sleep, including duration, quality, and mood. 
Standard question forms components such as multiple-choice, text input, sliders, and time pickers were utilized to ensure comprehensive data collection. 
Furthermore, interactive cognitive tasks, designed to resemble games, were employed to assess participants' cognitive skills.
The flexibility of ZotCare's capabilities in creating customized interaction components proved invaluable for researchers. 
This allowed them to focus on designing the tasks themselves without the need for developing a separate interactive mobile application. 
By leveraging ZotCare's functionalities, researchers could streamline the data collection process and efficiently collect data.
Given that some questionnaires needed to be administered before bedtime or upon waking up, ZotCare's profile and logic services were utilized to personalize the timing and availability of the questionnaires for each individual participant. 
This customization ensured that participants received questionnaires at the appropriate times based on their specific sleep schedules and preferences.

\begin{figure*}[ht]
\centering
\includegraphics[width=.8\textwidth]{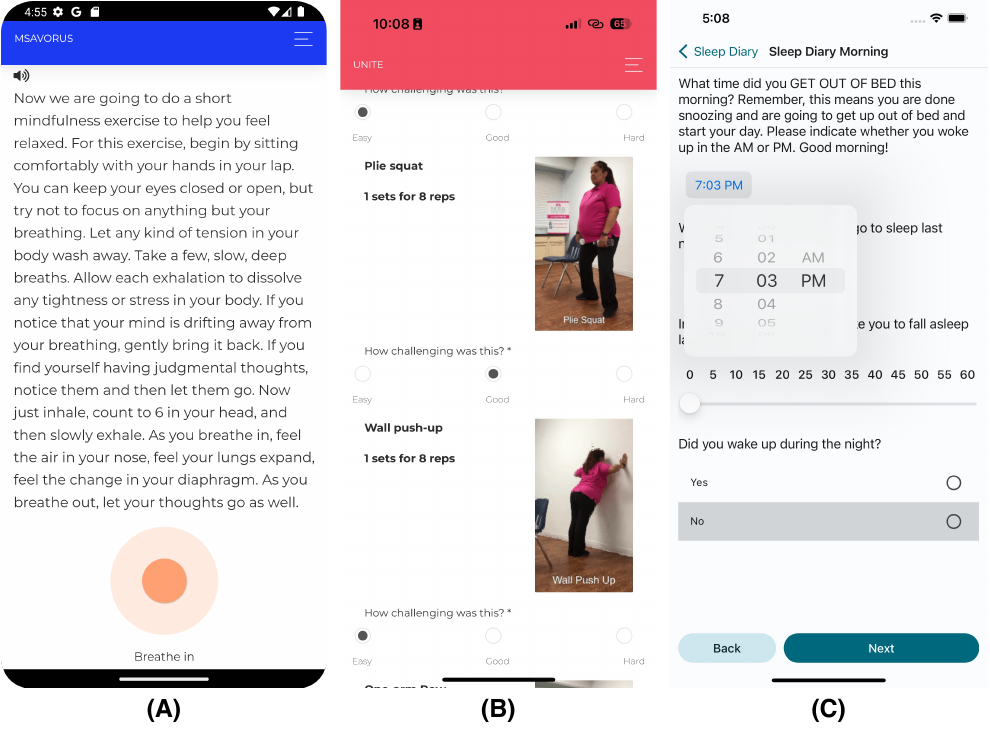}
\caption{(A) mSavorUs, (B) UNITE, and (C) HowRU app used in MHN, UNITE, and Brain \& Sleep projects }
\label{fig:unite-app}
\end{figure*}

In addition to its utilization in the aforementioned studies conducted in the United States, ZotCare has also been employed in research studies conducted in other countries and across different languages. 
One notable example is the ``PREVENT" study conducted in Finland. This study specifically focused on maternal care and aimed to assess the daily well-being of pregnant women during the challenging circumstances imposed by the COVID-19 pandemic \citep{finland-slim-study}.
The ``PREVENT" study leveraged ZotCare's localization features to adapt the platform to the Finnish language and timezone. 
This ensured that the study participants in Finland could access and interact with ZotCare's services seamlessly in their native language and within the context of their local time zone.

The Digital Health for the Future of Community-Centered Care (D-CCC) \citep{dccc-paper} research project aims to explore the integration of technology and community health workers in order to enhance healthcare delivery for underserved communities.
Specifically focusing on the elderly population, the project seeks to develop a symbiotic relationship between humans and technology, enabling the design of new technologies that can assist community health workers in providing more effective support.
ZotCare played a crucial role in alleviating the burden on participants who were unfamiliar with using advanced technologies. 
By providing user-friendly interfaces and intuitive interactions, ZotCare facilitated the seamless integration of technology into the participants' lives. 
Moreover, ZotCare enabled the continuous monitoring of participants' device usage patterns and their overall health status. 
This feature helped researchers and community health workers gain valuable insights into the participants' well-being and promptly address any emerging issues.
The ZotCare dashboard proved to be an invaluable tool in monitoring vital signs trends among participants, such as heart rate and blood pressure. 
Additionally, the ZotCare application allowed for the subjective capture of daily symptoms, including pain and fatigue, as well as adverse events such as falls.
The customization of the ZotCare dashboard specifically catered to the needs of health providers involved in this study. 
It enabled them to receive intelligent alerts in the event of abnormalities in vital signs or adverse events, and also provided visualizations of the collected data, empowering them to make informed decisions regarding participant care.

\section{Using ZotCare}

To acquire more information on how to use ZotCare, please visit \url{https://futurehealth.uci.edu/projects/zotcare/}.

\section{Limitations and Future Work}

This article primarily highlights the services and benefits of ZotCare in the context of mHealth research. 
However, it is essential to note that certain aspects, such as ZotCare's system architecture, technical contributions, and system challenges, have not been addressed explicitly in this article. 
These topics warrant separate and dedicated publications to provide in-depth insights and analysis. 

The future of ZotCare will predominantly hinge on two principal enhancements: the integration of new features and the introduction of novel services.
Additional features can be incorporated into the existing services without altering the current orchestration, for instance, augmenting support for new devices, introducing new types of profiles, or incorporating new actions and triggers into RPII services. 
Other relevant features, such as chatbots, could serve as beneficial additions to ZotCare's interactions.
Furthermore, there is potential for increasing the programmability of AI components within the system.
The system could be trained to suggest correlations and models congruent with researchers' experimental objectives, mitigating the need for manual programming.
Alterations in service orchestration would pose a more significant challenge, as the services must retain their flexibility, minimalism, and comprehensiveness. 
Nevertheless, with the burgeoning demand in the mHealth sector, the team will explore opportunities for redesigning the service orchestration.

While programmable services offer a high level of customization, they adhere to stringent syntax and rules, necessitating a learning phase for researchers. 
Besides that, some functionalities and patterns in mHealth systems often repeat in a similar way between studies, such as smart personalized notifications, standard data processing methods, and standard adaptive interventions.
In order to make it simpler for researchers, efforts are underway to incorporate more templates and straightforward, high-level configurations in the form of modules. 
These modules can be available to researchers to add standard mHealth functionalities to their studies without the need to learn knowledge over ZotCare programmable services or recreate standard functionalities that have been done before.

\section{Conclusion}

mHealth solutions are significant for researchers aiming to design studies that leverage internet-connected health devices and smartphones. 
These solutions must offer fast delivery, affordability, and the necessary programmability to support personalization and adaptation, particularly in the context of modern AI-driven investigations. 
To achieve this, we introduced ZotCare as a service-based solution that provides a ready-to-use platform for conducting mHealth studies while also enabling resource sharing to reduce costs. 
This paper primarily focused on ZotCare's service orchestration, features, usage, and capabilities.
ZotCare's service orchestration comprises Collection Services, Profile Services, and Real-time Processing, Integration, and Intervention (RPII) Services. Collection Services facilitate the aggregation of both objective and subjective data streams into the system, while Profile Services offer programmable key-value storage for participants. 
These services empower RPII Services to process data, generate models, and trigger tailored actions based on varying circumstances. 
Finally, we demonstrated the practical applications of ZotCare's services through various use cases. 
These examples indicated how different mHealth studies across diverse domains have successfully utilized ZotCare's programmability.

\section*{Acknowledgements}

This research was supported by the US National Science Foundation, under the Smart and Connected Communities (S\&CC) program (grant CNS-1831918).
We want to extend our sincere gratitude to Arman Anzanpour for his assistance with creating and refining the figures and graphics presented in this paper.

\bibliographystyle{Frontiers-Vancouver} 
\bibliography{main}





\end{document}